\documentclass[11pt,a4paper]{article}

\usepackage[T1]{fontenc}
\usepackage[utf8]{inputenc}
\usepackage{lmodern}
\usepackage[margin=2.5cm]{geometry}
\usepackage{amsmath,amssymb,amsthm,mathtools}
\usepackage{bm}
\usepackage{graphicx}
\usepackage{booktabs}
\usepackage{array}
\usepackage{enumitem}
\usepackage[hidelinks]{hyperref}

\newtheorem{proposition}{Proposition}
\newtheorem{corollary}{Corollary}
\theoremstyle{remark}
\newtheorem{remark}{Remark}
\theoremstyle{definition}
\newtheorem*{example*}{Example}

\newcommand{\Ind}{\mathcal{I}}
\newcommand{\Tr}{\operatorname{Tr}}
\newcommand{\Var}{\operatorname{Var}}
\newcommand{\supp}{\operatorname{supp}}
\newcommand{\E}{\mathbb{E}}
\newcommand{\ket}[1]{\lvert #1\rangle}
\newcommand{\bra}[1]{\langle #1\rvert}
\newcommand{\proj}[1]{\ket{#1}\!\bra{#1}}
\newcommand{\sig}{\bm{\sigma}}
\newcommand{\Pit}{\tilde{\Pi}}
\newcommand{\rhot}{\tilde{\rho}}

\title{Instrument indeterminacy: separating uncertainty in POVM effects\\ and its neutrosophic reading}
\author{Fernando A. Otero\\[4pt]
\small Departamento de Matemática, Facultad de Ingeniería,\\
\small Universidad Nacional de Mar del Plata, Mar del Plata, Argentina\\
\small \texttt{foterovega@fi.mdp.edu.ar}}
\date{}

\begin{document}
\maketitle

\begin{abstract}
Neutrosophic annotation frameworks for quantum measurement attach an external triple $(t,i,f)$ to each Born probability, and explicitly leave open how indeterminacy could be defined operationally. We show that indeterminacy cannot be recovered from outcome probabilities, but can be computed from the pair (state, effect). For a binary POVM $\{A, I-A\}$ we define the instrument indeterminacy $\Ind_\rho(A) = 4\Tr\bigl(\rho(A-A^2)\bigr)$, which vanishes exactly when the state only sees the sharp part of the effect. The normalized Gini index of the outcome splits as $4TF = \Ind_\rho(A) + 4\Var_\rho(A)$, a law of total variance that separates the instrument contribution from the state contribution; the identity is a special case of Massar's POVM uncertainty operator, and our contribution is its neutrosophic reading. Moving the system--apparatus cut to a Naimark dilation transfers exactly $\Ind_\rho(A)$ from indeterminacy to variance, which yields the missing operational definition. We extend the decomposition to $m$-outcome POVMs, relate the minimum indeterminacy of two incompatible unbiased qubit observables to their commutator through Busch's joint-measurability criterion, and show that $\Ind_\rho(A)$ equals twice the disagreement probability of a repeated Lüders measurement. The resulting estimator needs neither the state nor the effect and, in simulation, requires 2.5 to 6 times fewer shots than detector tomography for the same error.

\end{abstract}

\section{Introduction}

Neutrosophy describes a proposition by three independent degrees of truth, indeterminacy and falsity, $(t,i,f)$~\cite{smarandache1998}. Its transfer to quantum computing has produced neutrosophic states, gates and measurements~\cite{cevik2018,jafari2026}. These annotation frameworks work with records $(p_j,\nu_j)$: the probability $p_j = \Tr(\rho E_j)$ is fixed by quantum theory, whereas the triple $\nu_j = (t,i,f)$ is supplied from outside. The volume \emph{Neutrosophic Quantum Computing}~\cite{jafari2026} states this explicitly: indeterminacy is neither $p_j$, nor $1-p_j$, nor the variance of an observable (§6.10.2), and replacing $p_j$ by $(t,i,f)$ would require a separate theory (§6.11.2). The operational definition of indeterminacy appears there as an open problem (§6.12, Problem~1).

Randomness and indeterminacy are thus separated by stipulation. The reason is structural: at the level of probabilities, indeterminacy is invisible. Table~\ref{tab:same-stats} shows two situations with the same outcome statistics. In the first, all the uncertainty comes from the state; in the second, all of it comes from the instrument. No function of $(T,F)$ can tell them apart.

\begin{table}[ht]
\centering
\caption{Two measurements with the same statistics and a different source of uncertainty.}
\label{tab:same-stats}
\begin{tabular}{lcccc}
\toprule
Case & $T$ & $F$ & $\Ind_\rho$ & $4\Var_\rho$\\
\midrule
Projector of $\sigma_z$, equatorial state ($r_z = 0$) & 0.50 & 0.50 & 0.00 & 1.00\\
Trivial effect $I/2$, polarized state ($r_z = 0.8$) & 0.50 & 0.50 & 1.00 & 0.00\\
\bottomrule
\end{tabular}
\end{table}

One level down, at the pair $(\rho,A)$, the distinction is computable: indeterminacy is the unsharpness of the effect. The theory of unsharp observables has studied this notion for decades~\cite{busch1986,busch1996,busch1995,heinosaari2012}, and Massar introduced an uncertainty operator that measures the additional noise of a POVM relative to a von Neumann measurement~\cite{massar2007}. This work claims no novelty for that quantity. Its contributions are: (i) reading it as the $I$ degree of the neutrosophic triple and showing that it answers the open problem of~\cite{jafari2026}; (ii) the characterization of the decomposition as a law of total variance with an instrument/state attribution, and its dependence on the system--apparatus cut via the Naimark dilation; (iii) the extension to $m$-outcome POVMs, which covers the normalized neutrosophic mode $\{M_T, M_I, M_F\}$; (iv) the reading of Busch's joint-measurability criterion~\cite{busch1986} as a minimum price in indeterminacy; and (v) an estimation protocol by Lüders re-measurement that requires knowledge of neither the state nor the effect.

The rest of the paper is organized as follows. Section~\ref{sec:prelim} fixes the notation. Section~\ref{sec:main} presents the main results. Section~\ref{sec:estimation} deals with estimation from finite shots. Section~\ref{sec:examples} works through examples taken from~\cite{jafari2026}. Section~\ref{sec:discussion} discusses scope, nomenclature and limitations, and Section~\ref{sec:conclusions} concludes.

\section{Preliminaries and notation}\label{sec:prelim}

Let $\mathcal{H}$ be a finite-dimensional Hilbert space, $\rho$ a state ($\rho \ge 0$, $\Tr\rho = 1$) and $A$ an effect, that is, $0 \le A \le I$. The pair $\{A, I-A\}$ is a binary POVM. We define
\[
T = \Tr(\rho A), \qquad F = \Tr\bigl(\rho(I-A)\bigr) = 1-T,
\]
\[
\Ind_\rho(A) = 4\Tr\bigl(\rho(A-A^2)\bigr), \qquad \Var_\rho(A) = \Tr(\rho A^2) - \Tr(\rho A)^2.
\]
For the analysis we use the spectral decomposition of the effect and the Born weights of the state on each eigenspace:
\[
A = \sum_k a_k \Pi_k, \qquad a_k \in [0,1], \qquad w_k = \Tr(\rho\,\Pi_k).
\]
We write the instrument indeterminacy as $\Ind_\rho(A)$. The volume~\cite{jafari2026} uses $U(\nu)$ for an annotation functional (§13.7.5) that is unrelated to this quantity.

\section{Main results}\label{sec:main}

\subsection{Unsharpness of the effect}

\begin{proposition}\label{prop:unsharp}
For every effect $A$ and every state $\rho$:
\begin{enumerate}[label=\arabic*.]
\item $0 \le \Ind_\rho(A) \le 1$.
\item $\Ind_\rho(A) = 0$ if and only if $\supp(\rho) \subseteq \ker(A-A^2) = E_0(A) \oplus E_1(A)$, the subspace on which $A$ acts as a projector.
\item $\Ind_\rho(A) = 1$ if and only if $\supp(\rho) \subseteq E_{1/2}(A)$, the eigenspace of eigenvalue $1/2$.
\end{enumerate}
\end{proposition}

\begin{proof}
By the spectral decomposition,
\[
A - A^2 = A(I-A) = \sum_k a_k(1-a_k)\,\Pi_k, \qquad 0 \le a_k(1-a_k) \le \tfrac14,
\]
so that $0 \le A - A^2 \le I/4$ in the operator order. Since $\rho \ge 0$ and $\Tr\rho = 1$, (1) follows. For (2), let $X = A - A^2 \ge 0$. Then $\Tr(\rho X) = \Tr(X^{1/2}\rho X^{1/2})$, and the trace of a positive operator vanishes only if the operator is zero. Hence $\Tr(\rho X) = 0$ is equivalent to $\rho^{1/2}X^{1/2} = 0$, that is, $\supp(\rho) \subseteq \ker X$. The kernel of $X$ is the sum of the eigenspaces with $a_k \in \{0,1\}$. For (3), the same argument is applied to the identity $\tfrac14 I - (A - A^2) = (A - \tfrac12 I)^2 \ge 0$.
\end{proof}

\begin{remark}
The condition in (2) is on the support of $\rho$, not on $A$ over all of $\mathcal{H}$. An effect that is not a projector can have $\Ind_\rho(A) = 0$ if the state only sees its sharp part. Indeterminacy is a property of the pair $(\rho,A)$, not of the effect in isolation.
\end{remark}

\subsection{Decomposition of uncertainty}

\begin{proposition}\label{prop:decomp}
For every effect $A$ and every state $\rho$,
\[
4TF = \Ind_\rho(A) + 4\Var_\rho(A).
\]
The left-hand side is the Gini index (Tsallis entropy of order 2~\cite{tsallis1988}) of the binary outcome, normalized to $[0,1]$. Both terms are non-negative and also lie in $[0,1]$.
\end{proposition}

\begin{proof}
Since $F = 1 - T$,
\[
\Ind_\rho(A) + 4\Var_\rho(A) = 4\Tr(\rho A) - 4\Tr(\rho A^2) + 4\Tr(\rho A^2) - 4\Tr(\rho A)^2 = 4T(1-T). \qedhere
\]
\end{proof}

\begin{remark}[Relation to the literature]
If the value 1 is assigned to the effect $A$ and 0 to $I-A$, Massar's uncertainty operator~\cite{massar2007} reduces to $A - A^2$, and Proposition~\ref{prop:decomp} is the decomposition of the variance of the outcome into the variance of the average observable plus the intrinsic noise of the POVM. The identity is therefore not new; what is added here is its reading as a separation between indeterminacy (instrument) and randomness (state).
\end{remark}

\paragraph{Interpretation: law of total variance.} Every measurement with the effect $A$ reproduces the statistics of a two-stage process:
\begin{enumerate}[label=\arabic*.]
\item The state selects channel $k$ with probability $w_k = \Tr(\rho\,\Pi_k)$; this is the Born rule for the sharp observable that diagonalizes $A$.
\item Given $k$, the instrument produces outcome 1 with probability $a_k$, like a coin with that bias.
\end{enumerate}
Let $X \in \{0,1\}$ be the outcome and $K$ the channel. Then $P(X=1) = \sum_k w_k a_k = T$, and the law of total variance gives
\[
\underbrace{\Var(X)}_{TF}
= \underbrace{\E\bigl[\Var(X \mid K)\bigr]}_{\sum_k w_k a_k(1-a_k) \,=\, \Ind_\rho(A)/4}
+ \underbrace{\Var\bigl(\E[X \mid K]\bigr)}_{\sum_k w_k a_k^2 - T^2 \,=\, \Var_\rho(A)}.
\]
The first term is the average noise of the coin and is contributed by the instrument. The second is the spread of the biases across the channels occupied by the state and is contributed by the state. The partition is fixed by the spectrum of $A$; it is not chosen. The two-stage process reproduces the outcome probabilities, not necessarily the post-measurement state: different instruments with the same effect may leave different states (cf.~\cite{jafari2026}, Remark~6.9.2). Since the proposition involves only probabilities, this does not affect it.

\subsection{The Naimark cut}

Moving the system--apparatus cut to include the ancilla of a Naimark dilation~\cite{heinosaari2012} transfers exactly $\Ind_\rho(A)$ from indeterminacy to variance. The total uncertainty does not change; its attribution does.

Every effect admits a dilation with a one-qubit ancilla, $\mathcal{K} = \mathbb{C}^2$. The isometry
\[
V\ket{\psi} = \sqrt{I-A}\,\ket{\psi} \otimes \ket{0} + \sqrt{A}\,\ket{\psi} \otimes \ket{1}
\]
satisfies $V^\dagger V = I$ and $V^\dagger(I \otimes \proj{1})V = A$. Extending it to a unitary $U$ on $\mathcal{H} \otimes \mathcal{K}$ with $U(\ket{\psi}\otimes\ket{0}) = V\ket{\psi}$, one obtains the dilated projector and state
\[
\Pit = U^\dagger\bigl(I_{\mathcal{H}} \otimes \proj{1}\bigr)U, \qquad \rhot = \rho \otimes \proj{0}.
\]

\begin{corollary}\label{cor:naimark}
With this notation:
\begin{enumerate}[label=\arabic*.]
\item $\Tr(\rhot\,\Pit) = T$: the outcome has the same statistics in both descriptions.
\item $\Ind_{\rhot}(\Pit) = 0$: in the dilated description the measurement is sharp.
\item The indeterminacy of the system is the increase in variance when the cut is moved:
\[
\Ind_\rho(A) = 4\Var_{\rhot}(\Pit) - 4\Var_\rho(A).
\]
\end{enumerate}
\end{corollary}

\begin{proof}
(1) is $\Tr\bigl(\rho\,(I\otimes\bra{0})\,\Pit\,(I\otimes\ket{0})\bigr) = \Tr(\rho A)$. (2) follows from $\Pit^2 = \Pit$. For (3), Proposition~\ref{prop:decomp} applied on $\mathcal{H}\otimes\mathcal{K}$ gives $4T(1-T) = 4\Var_{\rhot}(\Pit)$, and applied on $\mathcal{H}$ gives $4T(1-T) = \Ind_\rho(A) + 4\Var_\rho(A)$. It suffices to subtract.
\end{proof}

\paragraph{Operational definition.} $\Ind_\rho(A)$ is the part of the outcome uncertainty that the chosen system--apparatus cut attributes to the instrument. This makes precise the informal reading according to which unsharpness is the information left in the ancilla when the measurement is refined to a PVM, and answers Problem~1 of~\cite[§6.12]{jafari2026}.

\subsection{\texorpdfstring{$m$}{m}-outcome POVMs}

The decomposition holds for any POVM, with the Gini index of the outcome distribution in place of $4TF$. This covers the normalized neutrosophic mode $\{M_T, M_I, M_F\}$~\cite[§11.5]{jafari2026}.

\begin{proposition}\label{prop:mpovm}
Let $\{M_k\}_{k=1}^m$ be a POVM and $p_k = \Tr(\rho M_k)$. Then
\[
1 - \sum_{k=1}^m p_k^2 = \sum_{k=1}^m \Tr\bigl(\rho(M_k - M_k^2)\bigr) + \sum_{k=1}^m \Var_\rho(M_k),
\]
and all terms are non-negative.
\end{proposition}

\begin{proof}
Since the other effects are positive and sum to $I - M_k$, each effect satisfies $0 \le M_k \le I$; by the argument of Proposition~\ref{prop:unsharp}, $M_k - M_k^2 \ge 0$. Summing,
\[
\sum_k \bigl(p_k - \Tr(\rho M_k^2)\bigr) + \sum_k \bigl(\Tr(\rho M_k^2) - p_k^2\bigr) = 1 - \sum_k p_k^2. \qedhere
\]
\end{proof}

Since $1 - \sum_k p_k^2 \le 1 - 1/m$, the instrument indeterminacy normalized to $[0,1]$ is
\[
\Ind_\rho(\{M_k\}) = \frac{m}{m-1}\sum_{k=1}^m \Tr\bigl(\rho(M_k - M_k^2)\bigr).
\]
For $m = 2$, with $M_1 = A$ and $M_2 = I - A$, one has $(I-A) - (I-A)^2 = A - A^2$ and $\Var_\rho(I-A) = \Var_\rho(A)$; the formula returns $\Ind_\rho(A)$ and Proposition~\ref{prop:mpovm} reduces to Proposition~\ref{prop:decomp}.

\begin{remark}[Two distinct indeterminacies]
In the POVM $\{M_T, M_I, M_F\}$, $p_I = \Tr(\rho M_I)$ is the Born probability of the outcome declared indeterminate, whereas $\Ind_\rho(\{M_k\})$ is the unsharpness of the instrument. They do not coincide: if $M_I$ is a projector, $p_I$ can equal 1 with $\Ind_\rho = 0$. We propose to call them \emph{declared indeterminacy} and \emph{instrument indeterminacy}, respectively.
\end{remark}

\subsection{Instruments and complementarity}

Two instruments with the same effect have the same indeterminacy; they can only be told apart when followed by another measurement, and then Proposition~\ref{prop:mpovm} applies to the sequential POVM. Incompatibility, on the other hand, imposes a minimum instrument indeterminacy that does not depend on the state.

\paragraph{Sequential instruments.} An instrument $\{\mathcal{E}_a\}$ has effects $E_a = \mathcal{E}_a^*(I)$~\cite{kraus1983}. Since $\Ind_\rho(E_a)$ depends only on the effects, two instruments with the same $E_a$ are indistinguishable at that level. Their difference, which lies in the post-measurement state, is only observable if another measurement $\{F_b\}$ follows. The full sequence is again a POVM,
\[
G_{ab} = \mathcal{E}_a^*(F_b), \qquad \sum_b G_{ab} = E_a,
\]
and Proposition~\ref{prop:mpovm} applies to $\{G_{ab}\}$ without the need for a new measure.

\begin{example*}
Let $E_+ = \tfrac12(I + 0.6\,\sigma_z)$, $\rho = \tfrac12(I + 0.7\,\sigma_x + 0.2\,\sigma_z)$, and let the probe $\{F_b\}$ be the measurement of $\sigma_x$. We compare the Lüders instrument~\cite{luders1951}, $\mathcal{E}_a(\rho) = \sqrt{E_a}\,\rho\,\sqrt{E_a}$, with a measure-and-prepare instrument, $\mathcal{E}_a(\rho) = \Tr(\rho E_a)\proj{0}$ (Table~\ref{tab:instruments}).
\end{example*}

\begin{table}[ht]
\centering
\caption{Same effect, different instrument: the difference only appears in the sequential POVM.}
\label{tab:instruments}
\begin{tabular}{lcccc}
\toprule
Instrument & $p(a)$ & $\Ind_\rho(E_+)$ & $p(a,b)$ & Sequential $\Ind_\rho$\\
\midrule
Lüders & $(0.56,\ 0.44)$ & 0.640 & $(0.42,\ 0.14,\ 0.36,\ 0.08)$ & 0.667\\
Measure-and-prepare & $(0.56,\ 0.44)$ & 0.640 & $(0.28,\ 0.28,\ 0.22,\ 0.22)$ & 0.880\\
\bottomrule
\end{tabular}
\end{table}

The measure-and-prepare instrument erases the information about $\sigma_x$, and the second measurement is reduced to a coin; this is why its sequential indeterminacy is larger. The value depends on the chosen probe; a probe-independent version (supremum or average over a tomographically complete set) remains open.

\paragraph{Complementarity.} Consider two unsharp, unbiased qubit observables, $E^a_\pm = \tfrac12(I \pm \mathbf{a}\cdot\sig)$ and $E^b_\pm = \tfrac12(I \pm \mathbf{b}\cdot\sig)$, with sharpnesses $\eta_a = |\mathbf{a}|$, $\eta_b = |\mathbf{b}|$ and angle $\theta$ between $\mathbf{a}$ and $\mathbf{b}$. By Example~1 (Section~\ref{sec:examples}), their indeterminacies are $\Ind_a = 1 - \eta_a^2$ and $\Ind_b = 1 - \eta_b^2$, independent of the state. Busch's criterion~\cite{busch1986} states that they are jointly measurable if and only if
\[
|\mathbf{a} + \mathbf{b}| + |\mathbf{a} - \mathbf{b}| \le 2.
\]

\begin{proposition}\label{prop:joint}
With this notation:
\begin{enumerate}[label=\arabic*.]
\item If $\mathbf{a} \perp \mathbf{b}$, the observables are jointly measurable if and only if $\Ind_a + \Ind_b \ge 1$.
\item If $\eta_a = \eta_b$, they are jointly measurable if and only if each indeterminacy satisfies
\[
\Ind_a = \Ind_b \ge \frac{\sin\theta}{1+\sin\theta}, \qquad \sin\theta = \tfrac12\bigl\lVert[\hat{\mathbf{a}}\cdot\sig,\ \hat{\mathbf{b}}\cdot\sig]\bigr\rVert.
\]
\end{enumerate}
\end{proposition}

\begin{proof}
(1) If $\mathbf{a} \perp \mathbf{b}$, $|\mathbf{a}+\mathbf{b}| = |\mathbf{a}-\mathbf{b}| = (\eta_a^2 + \eta_b^2)^{1/2}$ and the criterion becomes $\eta_a^2 + \eta_b^2 \le 1$, that is, $\Ind_a + \Ind_b \ge 1$. (2) If $\eta_a = \eta_b = \eta$, then $|\mathbf{a}+\mathbf{b}| = 2\eta\cos(\theta/2)$ and $|\mathbf{a}-\mathbf{b}| = 2\eta\sin(\theta/2)$. The criterion becomes $\eta\bigl(\cos(\theta/2) + \sin(\theta/2)\bigr) \le 1$ and, squaring, $\eta^2 \le 1/(1+\sin\theta)$. For unit vectors, $[\hat{\mathbf{a}}\cdot\sig,\ \hat{\mathbf{b}}\cdot\sig] = 2i(\hat{\mathbf{a}}\times\hat{\mathbf{b}})\cdot\sig$, whose operator norm is $2\sin\theta$.
\end{proof}

\begin{table}[ht]
\centering
\caption{Minimum indeterminacy for jointly measuring two unbiased observables of equal sharpness.}
\label{tab:joint}
\begin{tabular}{cccc}
\toprule
Angle $\theta$ & $\lVert[\hat{\mathbf{a}}\cdot\sig,\ \hat{\mathbf{b}}\cdot\sig]\rVert$ & Maximum $\eta$ & Minimum $\Ind_a + \Ind_b$\\
\midrule
$90^\circ$ & 2.000 & 0.7071 & 1.000\\
$60^\circ$ & 1.732 & 0.7321 & 0.928\\
$45^\circ$ & 1.414 & 0.7654 & 0.828\\
$30^\circ$ & 1.000 & 0.8165 & 0.667\\
\bottomrule
\end{tabular}
\end{table}

Jointly measuring two incompatible observables forces a minimum instrument indeterminacy, which grows with the commutator of their sharp versions and does not depend on the state. If they commute, the minimum is 0; at maximal incompatibility, it is $1/2$ for each. This gives $\Ind$ a second operational meaning: the price of compatibility. The joint-measurement criterion is due to Busch~\cite{busch1986}, and the relation between unsharpness and joint measurability was already pointed out by Busch and Shilladay~\cite{busch2002}; the contribution here is its expression in terms of $\Ind$. The result covers qubits with unbiased effects. For different sharpnesses and arbitrary angle the general criterion holds, which does not reduce to a function of $\Ind_a$ and $\Ind_b$ alone. Biased effects and dimension greater than 2 remain open.

\section{Estimation from finite shots}\label{sec:estimation}

$\Ind_\rho(A)$ can be measured without knowing either the state or the effect: it suffices to repeat the Lüders measurement and count the disagreements.

\begin{proposition}[Non-repeatability]\label{prop:repeat}
The effect $A$ is measured with the Lüders instrument, with operators $K_1 = \sqrt{A}$ and $K_0 = \sqrt{I-A}$, and $A$ is measured again on the resulting state. Then
\[
\Ind_\rho(A) = 2\,P(\text{the two outcomes differ}).
\]
\end{proposition}

\begin{proof}
$P(1,0) = \Tr\bigl(\sqrt{A}\,\rho\,\sqrt{A}\,(I-A)\bigr) = \Tr\bigl(\rho\,\sqrt{A}(I-A)\sqrt{A}\bigr) = \Tr\bigl(\rho(A-A^2)\bigr)$, because $\sqrt{A}$ commutes with $A$. The same computation gives $P(0,1) = \Tr\bigl(\rho(A-A^2)\bigr)$. The sum is $\Ind_\rho(A)/2$.
\end{proof}

$\Ind_\rho(A)$ is, then, the lack of repeatability of the measurement that least disturbs the system. A sharp effect is repeatable and gives $\Ind = 0$, which is the operational version of Proposition~\ref{prop:unsharp}(2). The identity requires the Lüders instrument: with the same effect as in the example of Table~\ref{tab:instruments}, measure-and-prepare $\ket{0}$ gives $2P(\text{disagreement}) = 0.928$, whereas $\Ind_\rho = 0.640$.

\paragraph{Estimation protocols.} We compare two protocols:
\begin{enumerate}[label=\arabic*.]
\item \emph{Lüders re-measurement.} With $k$ disagreements in $N$ repetitions, $\hat{\Ind} = 2k/N$. Since $k \sim \mathrm{Binomial}(N, \Ind/2)$, the estimator is unbiased, with standard error $\sqrt{\Ind(2-\Ind)/N}$. The confidence interval is the Wilson interval~\cite{wilson1927} for $\Ind/2$, multiplied by 2. It requires knowledge of neither $\rho$ nor $A$, but it does require a non-destructive measurement.
\item \emph{Detector tomography.} The detector is probed with the six Pauli eigenstates ($N/6$ shots each), $\hat{A}$ is reconstructed by linear inversion, projected onto $0 \le \hat{A} \le I$, and $\Ind_\rho(\hat{A})$ is computed. It works with destructive detectors, but requires knowledge of $\rho$ and calibrated probes. It was implemented only for one qubit.
\end{enumerate}

\paragraph{Results.} We simulated 4000 replicates per point, with $N$ between 100 and 100\,000, in five cases: A ($\eta = 0.9$, $\Ind = 0.19$), B ($\eta = 0.5$, $\Ind = 0.75$), C (projector, $\Ind = 0$), D (effect $I/2$, $\Ind = 1$) and E (detector with $s = 0.95$ and $r = 0.08$ in $d = 4$, $\Ind = 0.247$). Table~\ref{tab:bias} summarizes bias and coverage.

\begin{table}[ht]
\centering
\caption{Bias and coverage of the two protocols.}
\label{tab:bias}
\small
\begin{tabular}{ccccc}
\toprule
Case & $\Ind$ & \shortstack{Lüders bias,\\ $N = 100$} & \shortstack{Lüders 95\% coverage,\\ $N = 100\ /\ 3000\ /\ 100\,000$} & \shortstack{Tomography bias,\\ $N = 100\ /\ 100\,000$}\\
\midrule
A & 0.190 & $-1.0\times10^{-4}$ & 0.963 / 0.950 / 0.952 & $-0.034$ / $-1.1\times10^{-4}$\\
B & 0.750 & $+1.5\times10^{-3}$ & 0.931 / 0.944 / 0.952 & $-0.094$ / $+2.3\times10^{-6}$\\
C & 0 & 0 (exact) & trivial ($\hat{\Ind} = 0$ always) & $+0.042$ / $+0.002$\\
D & 1 & $-3.2\times10^{-3}$ & 0.946 / 0.955 / 0.956 & $-0.110$ / $-1.0\times10^{-4}$\\
E ($d = 4$) & 0.247 & $+1.3\times10^{-3}$ & 0.954 / 0.947 / 0.948 & not implemented\\
\bottomrule
\end{tabular}
\end{table}

\begin{figure}[ht]
\centering
\includegraphics[width=\linewidth]{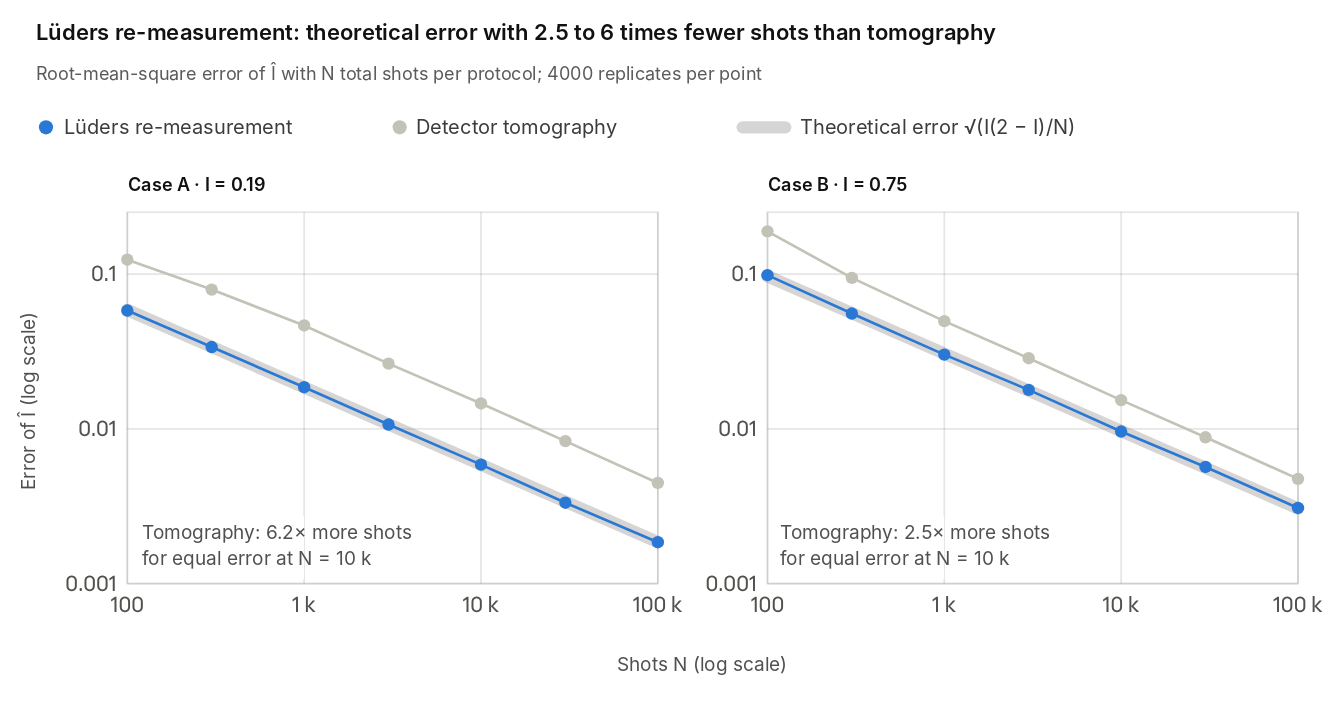}
\caption{Root-mean-square error of $\hat{\Ind}$ as a function of the total number of shots $N$ per protocol, cases A and B, 4000 replicates per point. The grey band is the theoretical error $\sqrt{\Ind(2-\Ind)/N}$.}
\label{fig:error}
\end{figure}

Lüders re-measurement attains the theoretical error in all cases, with slope $-0.500$ on a log-log scale (Figure~\ref{fig:error}). Tomography is biased at low shot counts. With a projector, the positive bias persists even at $N = 100\,000$, because the projection onto $[0, I]$ can only make the reconstructed effect unsharp. There are two special cases: with a projector, the re-measurement never disagrees and $\hat{\Ind} = 0$ is exact for any $N$; with $A = I/2$, the tomography error falls as $1/N$ (slope $-1.006$), because $\Ind$ is maximal there and the first-order error vanishes. To match the error of re-measurement, tomography needs between 2.5 and 6 times more shots; the advantage is larger when $\Ind$ is small.

\section{Examples}\label{sec:examples}

The first two examples are taken from~\cite{jafari2026}, where only probabilities are computed; here we additionally separate which part of the uncertainty is contributed by the instrument.

\paragraph{Example 1: unsharp measurement of a qubit (\cite{jafari2026}, §6.8.3).} Let $E_+ = \tfrac12(I + \eta\,\sigma_z)$ with $\eta \in [0,1]$, and $\rho = \tfrac12(I + \mathbf{r}\cdot\sig)$. The channels are $\ket{0}$ and $\ket{1}$, with biases $(1\pm\eta)/2$. Both coins have the same noise, $(1-\eta^2)/4$, so that
\[
\Ind_\rho(E_+) = 1 - \eta^2, \qquad 4\Var_\rho(E_+) = \eta^2(1 - r_z^2), \qquad 4TF = 1 - \eta^2 r_z^2.
\]
The indeterminacy depends only on the device; the state component depends on the polarization. The volume~\cite{jafari2026} only points out the case $\eta = 0$; Table~\ref{tab:qubit} shows how intermediate values are split.

\begin{table}[ht]
\centering
\caption{Splitting of the uncertainty in the unsharp measurement of a qubit.}
\label{tab:qubit}
\begin{tabular}{ccccc}
\toprule
$\eta$ & $r_z$ & $4TF$ & $\Ind$ (instrument) & $4\Var$ (state)\\
\midrule
0.5 & 0.0 & 1.00 & 0.75 & 0.25\\
0.9 & 0.0 & 1.00 & 0.19 & 0.81\\
0.5 & 0.8 & 0.84 & 0.75 & 0.09\\
\bottomrule
\end{tabular}
\end{table}

That $\Ind$ does not depend on the state has a consequence: an isotropic unsharpness ($A - A^2 \propto I$) does not distinguish states. To serve as a per-sample abstention criterion, the unsharpness must be concentrated on a subspace.

\paragraph{Example 2: detector with sensitivity $s$ and false-alarm rate $r$ (\cite{jafari2026}, §12.12.2).} The observed probability $p_{\mathrm{obs}} = sp + r(1-p)$ is $\Tr(\rho A)$ for the effect $A = sP + r(I-P)$. Its instrument indeterminacy is
\[
\Ind_\rho(A) = 4\bigl[s(1-s)\,p + r(1-r)(1-p)\bigr], \qquad p = \Tr(\rho P).
\]
With $s = 0.95$ and $r = 0.08$, for $p = 0.279$, $0.442$ and $0.556$ one obtains $\Ind = 0.265$, $0.248$ and $0.236$. This is the case in which unsharpness is the actual source of indeterminacy: noisy readout.

\paragraph{Example 3: same probabilities, different indeterminacy.} This is Table~\ref{tab:same-stats}. The projector of $\sigma_z$ on an equatorial state and the effect $I/2$ on a state with $r_z = 0.8$ both give $T = F = 0.50$, with $\Ind = 0$ and $\Ind = 1$, respectively. This makes precise the statement of~\cite[§1.3.4]{jafari2026} that $P(1) = 1/2$ does not imply $I = 1/2$.

\section{Discussion}\label{sec:discussion}

\paragraph{Dependence on the cut.} The separation is canonical once the system--apparatus cut is fixed. By Corollary~\ref{cor:naimark}, the total uncertainty $4TF$ is invariant under the cut, but its attribution is not. This is not a defect: it is the content of the operational definition. Any statement about $\Ind$ must declare what is taken as the system.

\paragraph{Names of the components.} The law of total variance underlying Proposition~\ref{prop:decomp} is the same one that separates uncertainties in evidential deep learning~\cite{sensoy2018,amini2020}. There, $\E[\Var]$ is called aleatoric and $\Var[\E]$ epistemic; here, $\E[\Var]$ corresponds to the instrument and $\Var[\E]$ to the state. To avoid confusion we call the terms the \emph{instrument component} ($\Ind$) and the \emph{state component} ($4\Var$).

\paragraph{Limitations.} The construction only describes the indeterminacy that arises in the measurement. Indeterminacy of statistical origin, for instance ambiguity in a ranking of methods, is outside its scope, and in preliminary explorations $\Ind$ did not improve on total uncertainty as a rejection criterion in that kind of task. In addition, the sequential version depends on the probe, and Proposition~\ref{prop:joint} is limited to qubits with unbiased effects.

\section{Conclusions}\label{sec:conclusions}

The neutrosophic indeterminacy of a quantum measurement, invisible in the outcome probabilities, becomes computable on the pair (state, effect) as the unsharpness $\Ind_\rho(A)$. Its natural complement is the variance of the observable, and both add up to the normalized Gini index of the outcome. The Naimark corollary gives it an operational definition: it is the part of the uncertainty that the chosen cut attributes to the instrument. The decomposition extends to $m$-outcome POVMs and distinguishes declared indeterminacy from instrument indeterminacy; incompatibility imposes a minimum of $\Ind$ that grows with the commutator; and Lüders re-measurement allows $\Ind$ to be estimated without knowing the state or the effect.

Future directions include a probe-independent version of sequential indeterminacy, the extension of Proposition~\ref{prop:joint} to biased effects and higher dimension, the implementation of detector tomography in arbitrary dimension, and the experimental validation of the re-measurement protocol on quantum hardware with non-destructive measurements.

\paragraph{Code availability.} All claims in Sections~\ref{sec:main} to~\ref{sec:examples} were verified numerically with a Python script (numpy and cvxpy), with fixed seeds, that reproduces the tables and Figure~\ref{fig:error}. The code is available upon request.

\paragraph{Funding.} This research received no external funding.

\paragraph{Conflicts of interest.} The author declares no conflicts of interest.

\end{document}